\documentclass[conference]{IEEEtran}
\IEEEoverridecommandlockouts
\usepackage{cite}
\usepackage{amsmath,amssymb,amsfonts}
\usepackage{algorithmic}
\usepackage{graphicx}
\usepackage{subcaption}
\usepackage{textcomp}
\usepackage{xcolor}
\usepackage{multirow}
\usepackage{hyperref}

\usepackage{adjustbox}
\usepackage{color}
\usepackage[normalem]{ulem}
\usepackage{comment}
\usepackage{todonotes}

\def\BibTeX{{\rm B\kern-.05em{\sc i\kern-.025em b}\kern-.08em
    T\kern-.1667em\lower.7ex\hbox{E}\kern-.125emX}}
    
\def\JNdel#1{\bgroup\markoverwith{\textcolor{red}{\rule[0.5ex]{2pt}{1pt}}}\ULon{#1}}

\begin{document}

\title{Cost-effective Models for \\Detecting Depression from Speech}
 \author{\IEEEauthorblockN{1\textsuperscript{st} Mashrura Tasnim}
 \IEEEauthorblockA{\textit{Winterlight Labs} \\
 Toronto, Canada \\
 mashrura@winterlightlabs.com}
 \and
 \IEEEauthorblockN{2\textsuperscript{nd} Jekaterina Novikova}
 \IEEEauthorblockA{\textit{Winterlight Labs} \\
 Toronto, Canada \\
 jekaterina@winterlightlabs.com}}

\maketitle

\begin{abstract}
Depression is the most common psychological disorder and is considered as a leading cause of disability and suicide worldwide. An automated system capable of detecting signs of depression in human speech can contribute to ensuring timely and effective mental health care for individuals suffering from the disorder. Developing such automated system requires accurate machine learning models, capable of capturing signs of depression. However, state-of-the-art  models based on deep acoustic representations require abundant data, meticulous selection of features, and rigorous training; the procedure involves enormous computational resources. In this work, we explore the effectiveness of two different acoustic feature groups - conventional hand-curated and deep representation features, for predicting the severity of depression from speech. We explore the relevance of possible contributing factors to the models' performance, including gender of the individual, severity of the disorder, content and length of speech. Our findings suggest that models trained on conventional acoustic features perform equally well or better than the ones trained on deep representation features at significantly lower computational cost, irrespective of other factors, e.g. content and length of speech, gender of the speaker and severity of the disorder. This makes such models a better fit for deployment where availability of computational resources is restricted, such as real time depression monitoring applications in smart devices.
\end{abstract}


\section{Introduction}
\label{sec:intro}
Depression is a common psychological disorder. About 264 million people worldwide suffer from depression, which is almost 5\% of the world's total population \cite{WHO}. Only about 50\% of the people experiencing major depression receive treatment. Due to lack of continuous monitoring and timely support, depression causes one death every 40 seconds, resulting in 800,000 deaths by suicide worldwide every year \cite{WHO}.

Conventional depression diagnostic systems, such as clinical assessment or standard questionnaires, require significant amount of time and active participation of the depressed individuals. Studies reveal that depression is reflected in behavioral fluctuations of certain day-to-day activities and physical parameters \cite{wang2014studentlife}. These findings have accelerated interventions in depression recognition using predictive models that incorporate input data of different modalities, among which audiovisual is one of the most explored areas. In this work, we emphasize on audio modality for its manifold benefits. Audio based depression detection system offers better privacy for users of remote monitoring system. This kind of automated assessment takes only a few minutes of audio recording, therefore is less time-consuming, and would reduce burden on the individuals.

Multiple research efforts aim to develop a system that detects depression by analyzing the fluctuation of acoustic features in human speech (\cite{ma2016depaudionet}, \cite{tasnim2019detecting}). An ML model that detects evidences of depression from audio data is a prerequisite for such a system. Existing best performing ML models that detect mental and cognitive diseases from audio data use either deep representation acoustic features, or a combination of conventional hand-crafted and deep features \cite{ray2019multi,diep2022multimodal}. Although deep representation features offer a unified process of feature extraction, feature selection and model training, extracting and processing these features demands enormous computation resources including memory and processing time. This makes such models inconvenient for many real-world applications, where speed of data processing, model training and inference are of crucial importance \cite{yalamanchili2020real}. Therefore, researchers and system designers need to make a choice of features when developing and deploying the model, considering both performance and cost. Some previous research compare the two approaches in the domain of cognitive disease detection \cite{balagopalan21_interspeech} but to the best of our knowledge, no such research has been done so far in the domain of depression. To address this gap, in this work we have experimented with both hand-crafted conventional acoustic features and deep representation acoustic features. 
We address the following research questions:
\begin{enumerate}
    \setlength{\itemsep}{1pt}
    \setlength{\parskip}{0pt}
    \setlength{\parsep}{0pt}
    \item \textit{Between conventional and deep representation acoustic features, which ones are more effective in determining depression severity in terms of accuracy and computational cost?}
    \item \textit{Does the machine learning (ML) model performance vary based on gender of the subject?}
    \item \textit{What is the effect of content and length of speech data in predicting depression from speech?}
\end{enumerate}
Answers to these questions enables the research community as well as system designers in making informed choice of modality, features, algorithm that suits best to the context, e.g. target user group and affordability. In this work, we compare performance of the ML models trained on each type of features extracted from speech samples of a variety of content and length. Our key findings suggest that: 
\begin{enumerate}    
    \setlength{\itemsep}{1pt}
    \setlength{\parskip}{0pt}
    \setlength{\parsep}{0pt}
    \item  ML model trained on conventional acoustic feature set curated using expert domain knowledge demonstrates competitive performance as state-of-the-art models in predicting depression severity, irrespective of length and content of speech, and gender of the speaker
    \item Usage of deep representation features resulted in marginal improvement of performance (0.0004\%) consuming 1000 times more memory and 3000 times more computation time.
\end{enumerate}
As such, we claim that models predicting depression from human speech that are trained on conventional acoustic features are a better choice than the models trained on deep acoustic representations in the situations when computational resources are limited, e.g. in mental healthcare applications for portable or wearable devices. On the other hand, deep representation models fit better to the scenarios where abundant training data is available, for example social media, and computational resources is a legitimate trade-off for better performance.

\section{Related Works}
\label{sec: background}
Individuals suffering from psychological and neurological disorders like depression exhibit measurable fluctuation in vocal parameters  (\cite{darby1984speech} and \cite{cummins2014probabilistic}). Significant number of research have been conducted to relate these parameters with depression severity. DAIC-WoZ dataset \cite{gratch2014distress} is a widely used dataset in acoustic based depression severity prediction, consisting of structured interviews of participants conducted by a virtual agent. Two subsets of this dataset have been introduced as the challenge corpus of three Audio/Visual Emotion Challenges (AVEC) in 2016 \cite{valstar2016avec}, 2017 \cite{ringeval2017avec} and 2019 \cite{ringeval2019avec}, where participants proposed machine learning models to predict depression score on the PHQ-8 scale \cite{kroenke2001phq}. Handcrafted acoustic features have been exploited for this task for the last few decades, while deep representation of acoustic features have become popular in recent years. Further, we present a summary of existing works in this area and compare them based on the type of acoustic features.

\subsection{Conventional Acoustic Features}
Conventional acoustic features fall in temporal, spectral, energy and voicing related categories, from which researchers hand-pick the ones that are most suitable for predicting certain disorders, such as depression \cite{cummins2014probabilistic}.
Over the time, certain sets of these acoustic features, introduced in speech emotion and depression recognition challenges, have gained popularity, among which baseline feature sets of AVEC 2013 \cite{valstar2013avec} and AVEC 2016 \cite{valstar2016avec}, INTERSPEECH ComParE \cite{schuller2013interspeech}, extended Geneva Minimalistic Acoustic Parameter Set (eGeMAPS) \cite{eyben2015real} are noteworthy. 
Development of feature extraction toolkits like openSMILE \cite{eyben2010opensmile}, COVAREP \cite{degottex2014covarep} has made it easier for researchers to extract these features for the purpose of speech analysis in different aspects.

\subsection{Deep Representation Features}
Deep representation of acoustic features are inspired by the deep learning paradigms common in image processing. Here, spectral images of speech instances are fed into pre-trained image recognition CNNs and a set of the resulting activations are extracted as feature vectors. In AVEC 2019 Depression Detection Sub-challenge (DSC), Deep representation features from four robust pre-trained CNNs using VGG-16 \cite{huang2017densely},
AlexNet \cite{krizhevsky2012imagenet}, DenseNet-121, and DenseNet-201 \cite{simonyan2014very} were included as challenge baseline features. Participants chose between using one or more sets of deep representation features (\cite{rodrigues2019multimodal}, \cite{yin2019multi}), and combining them with traditional features \cite{ray2019multi} and obtained competitive performances (Table \ref{tab:comparison}). Deep representation provides the option to unite feature extraction, feature selection and model training into a single automated generalisable procedure, compromising the opportunity to incorporate expert domain knowledge~\cite{novikova2022deck}, and necessitating considerably higher computational cost.

\subsection{Depression Detection Models}
AVEC 2016 challenge dataset was used in analysis presented in \cite{williamson2016detecting}, \cite{sun2017random}, \cite{gong2017topic}, \cite{yang2017hybrid}, \cite{samareh2018predicting}, \cite{al2018detecting}, \cite{haque2018measuring}, \cite{zhao2020hierarchical} and \cite{muzammel2020audvowelconsnet}. Williamson \textit{et al.} extracted formants, MFCCs, glottal features, loudness \cite{williamson2016detecting}. In addition to COVAREP \cite{degottex2014covarep} audio features, \cite{sun2017random} took text topics into account, while \cite{gong2017topic} considered a more extended set of features of audio, video and text modalities. \cite{samareh2018predicting} added Delta and Delta-Delta coefficients, mean, median, standard deviation, peak-magnitude to RMS ratio to the set of challenge baseline audio features. Applying similar higher-order statistics on the baseline features, \cite{al2018detecting} constructed an extended feature set of 553 features, of which they identified 279 features with statistically significant univariate correlation. \cite{haque2018measuring} implemented multi-modal sentence-level embedding on log-Mel spectrogram and MFCC features. In their recent work, \cite{yang2020feature} exploited a combination of eGeMAPS and INTERSPEECH features extracted from the longest ten segments of each audio sample. They reshape the feature vector in an image-like 2D feature map in row-major order and adopted Deep Convolutional Generative Adversarial Net (DCGAN) for feature vector generation. \cite{muzammel2020audvowelconsnet} trained three spectrogram-based Deep Neural Network architectures phoneme consonant and vowel units and their fusion. Their findings suggest that deep learned consonant-based acoustic characteristics lead to better recognition results than vowel-based ones, and the fusion of vowel and consonant speech characteristics outperforms the other models on the task. \cite{zhao2020hierarchical} described a transfer attention mechanisms from speech recognition to aid depression severity measurement. The transfer is applied in a two-level hierarchical network which reflecting the natural hierarchical structure of speech.

The AVEC 2016 challenge corpus included training, development and test partitions of audio samples. Using acoustic features exclusively, the lowest root-mean-square-error (RMSE) of 5.52 and 6.42 on the development and test set were reported in \cite{yang2017dcnn} and \cite{syed2017depression} respectively. RMSE 4.99, 5.40, 5.66 and 6.42 were reported in \cite{gong2017topic}, \cite{yang2017dcnn}, \cite{zhao2020hierarchical} and \cite{syed2017depression} respectively on the test set. The challenge baseline RMSE 6.74 (mean absolute error (MAE) 5.36) and 7.78 (MAE 5.72) were set for the development and test set respectively  \cite{ringeval2017avec}. 

Comparatively, fewer number of studies have been conducted on the recently published AVEC 2019 DDS dataset; which is a super-set of the AVEC 2016 dataset. 
A wide range of audio features has been provided as the challenge baseline encompassing both handcrafted sets of conventional features and deep representation of acoustic features, including eGeMAPS, Mel Frequency Cepstral Coefficients (MFCCs), Bag of Audio Words (BoAW) and two sets of deep spectrum features by feeding spectral images of speech instances into pre-trained image recognition Convolutional Neural Networks (CNN) (VGG-16 \cite{simonyan2014very} and DenseNet-121 \cite{huang2017densely}) and extracting the resulting activations as feature vectors. Of these, deep spectrum features were used by \cite{yin2019multi} and \cite{rodrigues2019multimodal}, and MFCCs and eGeMAPS by \cite{fan2019multi}. \cite{ray2019multi} exploited all the baseline feature sets, while \cite{zhang2020automated} extracted the AVEC 2017 baseline feature set \cite{ringeval2017avec} using the COVAREP software toolbox \cite{degottex2014covarep}, in addition to eGeMAPS. Different configurations of long short-term memory (LSTM) networks were used in all of these works except Zhang \textit{et al.} \cite{zhang2020automated}, who adopted random forest and logistic regression. Acoustic models proposed by \cite{ray2019multi} and \cite{zhang2020automated} achieved lowest RMSE of 5.11 and 6.78 on the development and test partitions, respectively.

\section{Materials and Methods}
\label{sec: procedure}

\subsection{Dataset}
We use DEPression and Anxiety Crowdsourced corpus (DEPAC) \cite{tasnim2022DEPAC} in this experiment. The dataset consists of 2,674 audio samples collected from 571 subjects located in Canada and the United States. 54.67\% of the study subjects are female and 45.33\% are male, aged between 18 and 76 years, and they received 1 to 26 years of formal education. The data was collected via crowdsourcing and consists of a variety of self-administered speech tasks (Table \ref{tab:speech_task}). The participants completed these tasks using Amazon Mechanical Turk (mTurk) \footnote{\url{https://www.mturk.com}}. The speech tasks were curated to increase phonemic variety and were supported by literature on detecting mental disorders, such as Alzheimer’s Disease (AD) \cite{borkowski1967word} and depression \cite{jiang2017investigation}, \cite{fossati2003qualitative}, \cite{trifu2017linguistic} from speech. 

\begin{table}[h]
\vspace{-1em}
\renewcommand{\arraystretch}{1.3}
    \centering
    \caption{Speech tasks in DEPAC corpus}
    \begin{tabular}{|p {0.15\linewidth}|p{0.5\linewidth}|p {0.15\linewidth}|}
    \hline
    \textbf{Speech Task} & \textbf{Description} & \textbf{Average Duration} \\
    \hline
    Phoneme Task & Record “aah” sound for as long as the participant could hold breath & 5.79 sec\\
    \hline
    Phonemic Fluency & Pronounce as many unique words as possible starting with the letters “F”, “A” or “S” & 22.13 sec\\
    \hline
    Picture Description & Describe a picture shown on the screen & 46.60 sec \\
    \hline
    Semantic Fluency & Describe a positive experience they expected to have within five years in future & 43.76 sec\\
    \hline
    Prompted Narrative & Tell a personal story, describing the day, a hobby, or a travel experience & 45.34 sec\\
    \hline
    \end{tabular}
    \label{tab:speech_task}
\end{table}

In this dataset, the depression severity is represented by Patient Health Questionnaire (PHQ-9) scores. It is a 3-point self-rated measure for depressive symptoms, including 9 questions. To ensure comparability of our results with works done on popular subsets of DAIC-WoZ corpus \cite{gratch2014distress}, i.e. AVEC 2017 \cite{ringeval2017avec} and AVEC 2019 \cite{ringeval2019avec}, we used responses to 8 PHQ questions in our analysis and reported our results on PHQ-8 scores. The score ranges from 0 to 24 on PHQ-8 scale where a score higher than 5, 9 and 14 represent mild, moderate and severe level of depression respectively. The mean PHQ-8 score of DEPAC corpus (M) is 6.56 with standard deviation (SD) of 5.56.

\subsection{Audio Quality Enhancement}
To suppress possible background noise present in the samples and improve quality of the audio, we applied \textit{logmmse} enhancement technique \cite{ephraim1985speech} on the audio samples. This method was found the best among existing audio enhancement algorithms \cite{hu2006subjective}. The enhancement step is found statistically significant $(p \leq 0.005)$ on 94\% of the 220 conventional acoustic features in Wilcoxon signed-rank test with Bonferroni correction.

Audio volume was normalized to -20 dBFS across all speech segments to control for variation caused by recording conditions such as microphone placement.

\subsection{Acoustic Features}
We extracted two sets of acoustic features, representing hand-crafted sets of conventional features and deep learning features:
\subsubsection{Conventional acoustic features}
This set included 220 acoustic features, extracted from each audio sample. The feature set includes:

 \begin{itemize}
  \setlength{\itemsep}{1pt}
  \setlength{\parskip}{0pt}
  \setlength{\parsep}{0pt}
    \item \textbf{Spectral features: } Intensity (auditory model based), MFCC 0-12, Zero-Crossing Rate (ZCR)
    \item \textbf{Voicing-related features: }Fundamental frequency $(F_0)$, Harmonic-to-Noise Ratio (HNR), shimmer and jitter, durational features, pauses and fillers, phonation rate
\end{itemize}
Statistical functionals including minimum, maximum, average, and variance were computed on the low-level descriptors. Additionally, skewness and kurtosis were calculated on MFCCs, first and second order derivatives of MFCCs, and Zero Crossing Rate (ZCR) \cite{low2020automated}.

A Python implementation of Praat phonetic analysis toolkit \cite{boersma2001speak} has been used to extract the majority of these features. The MFCC features and their functionals were computed using \verb|python_speech_features|\footnote{\url{https://pypi.org/project/python\_speech\_features/}} library.

\subsubsection{Deep Representation Features}
Deep representation of acoustic features are inspired by the deep learning paradigms common in image processing. Here, spectral images of speech instances are fed into pre-trained image recognition CNNs and a set of the resulting activations are extracted as feature vectors. VGG-16 is a type of Convolutional Neural Network (CNN) which is considered to be one of the best computer vision models to date. We used DeepSpectrum library \cite{amiriparian2017Snore} to extract features from a pre-trained VGG-16 CNN \cite{simonyan2014very}. The speech files are first transformed into mel-spectrogram images with 128 mel-frequency bands. Then, the spectral images are forwarded through the pre-trained networks. A 4,096-dimensional feature vector is then formed from the activations of the second fully connected layer in VGG-16. The features were extracted at a window width of 1s and a hop size of 300 ms from each audio sample. 

\subsection{Data Preprocessing}
\subsubsection{Standardization }
The range of values of audio features tends to vary widely.
To ensure even contribution of all features in the regression task, and to speed up gradient descent convergence of the deep neural network, once acoustic features were extracted from the audio samples, we standardized them using z-scores, i.e., subtracting the mean and dividing by standard deviation. The standard score of a sample $x$ of feature $f_i$ is calculated as:
\begin{equation}
y = \frac{x-\mu}{\sigma}
\end{equation}
here $\mu$ and $\sigma$ are the mean and standard deviation of the values of $f_i$ in all training samples.

\subsubsection{Feature Selection}
We applied minimum Redundancy-Maximum Relevance (mRMR) algorithm \cite{zhao2019maximum} to select the most relevant features with respect to the PHQ scores, minimizing redundancy in the selected set of features. 10\% features were selected from each set for training the ML models. 

\subsection{Model Training}
Following \cite{balagopalan21_interspeech} and \cite{tasnim2019detecting}, we train a combination of linear and non-linear ML models separately on conventional and deep learning acoustic features:  

 \begin{itemize}
  \setlength{\itemsep}{1pt}
  \setlength{\parskip}{0pt}
  \setlength{\parsep}{0pt}
  \item Support Vector Machines (SVM): Radial Basis Function (RBF) kernel SVM was trained. Values of hyperparameters \textit{`C'} and \textit{`gamma'} were tuned by 5-fold grid-search cross validation (cv).
  
  \item Random Forest (RF): Scikit Learn implementation of random forest regressor was used. Number of estimator trees and maximum depth were tuned through grid-search cv.
  
  \item Feedforward Neural Network (FNN): The FNN model consists of 4 hidden layers, with 500 hidden units on the first layer, 250 in the second and 125 in the rest of the hidden layers. 30\% dropout on the output of each of the hidden layers of the FNN. We use Adam optimizer in all FNN models with the learning rate of 0.001. Each of the FNN models is trained for 150 epochs.
\end{itemize}

The discussion presented by \cite{balagopalan21_interspeech} suggest that for small audio corpus like the ADReSSo challenge dataset (237 samples) \cite{luz2021detecting}, either leave-one-subject-out cross validation or k-fold cross validation can be applied. However, the dataset used is this work is considerably larger than the ADReSSo challenge dataset. Considering the size of the dataset and corresponding computational complexity, we decided to report evaluation metrics with  5-fold cross-validation (CV) for the models. We create 5 subject-independent folds, train the model using 4 of them, and use the rest for testing. We repeat the process for all 5 folds and report evaluation metrics averaging across predictions on all the folds. These folds preserve the same ratio of depression severity in each training and test partitions.

To understand the effect of speech content on ML models' performance, we separated samples with each type of speech task and trained models on each type of them. We repeated the same process for conventional and VGG-16 features.

We ran our experiment on a MacBook Pro with Intel Core i7 CPU at clock speed of 2.67 GHz. The system availed 16 GB memory. The data preprocessing and model training was done in Python programming language.

\begin{table}[h]
\vspace{-1em}
\renewcommand{\arraystretch}{1.3}
    \centering
    \caption{Regression error of models trained on conventional and VGG-16 features}
    \setlength\tabcolsep{2pt}
    \begin{tabular}{|p {0.15\linewidth}|p{0.12\linewidth}|r|r|r|r|}
    \hline
    \multirow{2}{*}{\textbf{Algorithm}} & \multirow{2}{*}{\textbf{Gender}} & \multicolumn{2}{|c|}{\textbf{RMSE}} & \multicolumn{2}{|c|}{\textbf{MAE}}\\
    \cline{3-6}
    & & Conventional & VGG-16 & Conventional & VGG-16\\
    \hline
    \multirow{3}{*}{SVM} & Male & 5.04 & 7.89 & 4.22 & 6.95\\
    & Female & 5.64 & 7.11 & 4.33 & 6.23 \\
    & \textbf{Overall} & 5.38 & 7.48 & 4.28 & 6.56 \\
    \hline
    \multirow{3}{*}{RF} & Male & 5.15 & 5.06 & 4.37 & 4.27\\
    & Female & 5.47 & 5.51 & 4.34 & 4.32 \\
    & \textbf{Overall} & 5.32 & 5.31 & 4.31 & 4.33\\
    \hline
    \multirow{3}{*}{FNN} & Male & 5.10 & 5.19 & 4.45 & 4.30\\
    & Female & 5.54 & 5.67 & 4.51 & 4.34 \\
    & \textbf{Overall} & 5.35 & 5.46 & 4.40 & 4.32\\
    \hline
    \end{tabular}
    \label{tab:overall_performance}
\end{table}

\begin{table}[ht]
    \caption{Comparison of performance of SOTA ML models trained on different combinations of features. Bold denotes regression error of our proposed model.}
\renewcommand{\arraystretch}{1.3}
    \centering
    \begin{tabular}{|p{0.15\linewidth}|p{0.4\linewidth}|r|r|}
        \hline
        \textbf{Feature Type} & \textbf{Study} & \textbf{RMSE} & \textbf{MAE} \\ 
        \hline
        \multirow{7}{5em}{Conventional} &  Formants, MFCCs, glottal features, loudness, AVEC 2017 dataset \cite{williamson2013vocal} & 6.38 & 5.32\\
        \cline{2-4}
        & COVAREP feature set, AVEC 2017 dataset \cite{syed2017depression}  & 6.34 & 5.30 \\ 
        \cline{2-4}
        &  COVAREP features and functional, AVEC 2016 dataset) & 6.50 & 5.13 \cite{al2018detecting}\\
        \cline{2-4}
        &  MFCC, AVEC 2016 dataset \cite{haque2018measuring} & 5.78 & - \\
        \cline{2-4}
        &  MFCC and eGeMAPS features, AVEC 2019 dataset \cite{fan2019multi} & 6.20 & - \\ 
        \cline{2-4}
        &  eGeMAPS, INTERSPEECH features, AVEC 2016 dataset \cite{yang2020feature} & 5.52 & 4.63 \\
        \cline{2-4}
        &  eGeMAPS and COVAREP features, AVEC 2019 dataset \cite{zhang2020automated} & 6.78 & 5.77\\
        \hline
        \multirow{2}{5em}{Deep representation} & VGG-16 features, AVEC 2019 dataset \cite{rodrigues2019multimodal}  & 5.70 & - \\
        \cline{2-4}
        &  Mel-spectra, AVEC 2017 dataset \cite{zhao2020hierarchical} & 5.66 & 4.28 \\
        \hline
        Conventional + deep combined &  MFCC, BoAW, eGeMAPS and VGG-16 features, AVEC 2019 dataset \cite{ray2019multi} & 5.11 & - \\
        \noalign{\hrule height 2pt}
        Conventional & MFCCs, HNR, jitter, shimmer, ZCR features, DEPAC dataset \cite{tasnim2022DEPAC} & \textbf{5.31} & \textbf{4.33} \\
        \hline
    \end{tabular}
    \label{tab:comparison}
\end{table}

\begin{table}[ht]
\vspace{-1em}
\renewcommand{\arraystretch}{1.3}
    \centering
    \caption{Time elapsed in different stages of model training}
    \setlength\tabcolsep{2pt}
    \begin{tabular}{|p{0.2\linewidth}|l|r|r|}
    \hline
    \textbf{Processing step}& \textbf{Algorithm} & \textbf{Conventional} & \textbf{VGG-16}\\
    \hline
    Data loading & - & 1.132 & 244450\\
    \hline
    Preprocessing & - & 96.834 & 545483\\
    \hline
    \multirow{3}{*}{Model training} & SVM & 1.600 & 5593\\
    & RF & 0.715 & 981\\
    & FNN & 220.853 & 10270\\
    \hline
    \multirow{3}{*}{Prediction} & SVM & 0.412 & 53 \\
    & RF & 0.040 & 9\\
    & FNN &  1.102 & 7 \\
    \hline
    \multirow{3}{*}{\textbf{Total}} & SVM & 99.978 & 795579\\
    & RF & 98.715 & 790923\\
    & FNN & 267.227 & 800210\\
    \hline
    \end{tabular}
    \label{tab:elapsed_time}
\end{table}

\begin{table}[ht]
\vspace{-1em}
\renewcommand{\arraystretch}{1.3}
    \centering
    \caption{Regression error of models trained on speech samples of different tasks}
    \setlength\tabcolsep{2pt}
    \begin{tabular}{|l|p{0.17\linewidth}|r|r|r|r|}
    \hline
    \multirow{2}{*}{\textbf{Algorithm}} & \multirow{2}{*}{\textbf{Speech task}} & \multicolumn{2}{|c|}{\textbf{RMSE}} & \multicolumn{2}{|c|}{\textbf{MAE}}\\
    \cline{3-6}
    & & Conventional & VGG-16 & Conventional & VGG-16\\
    \hline
    \multirow{5}{*}{SVM} & Semantic fluency & 5.30 & 6.62 &4.21 & 5.77 \\
    & Prompted narrative & 5.29 & 6.63 & 4.24 & 5.77 \\
    & Phoneme task &  5.49 & 6.40 & 4.33 & 5.54 \\
    & Phonemic fluency & 5.45 & 6.49 & 4.35 & 5.62 \\
    & Picture description & 5.43 & 6.54  & 4.36 & 5.67 \\
    \hline
    \multirow{5}{*}{RF} & Semantic fluency & 5.24 & 5.24 & 4.25 & 4.24 \\
    & Prompted narrative & 5.31 & 5.25 & 4.30 & 4.25 \\
    & Phoneme task &  5.39 & 5.29  & 4.37 & 4.30 \\
    & Phonemic fluency & 5.38 & 5.29  & 4.38 & 4.30\\
    & Picture description & 5.42 & 5.31  & 4.40 & 4.31\\
    \hline
    \multirow{5}{*}{FNN} & Semantic fluency & 5.34 & 7.13 & 4.28  & 5.34 \\
    & Prompted narrative & 5.30 & 7.13 & 4.30 & 5.35 \\
    & Phoneme task &  5.49 & 7.33  & 4.39 & 5.50\\
    & Phonemic fluency & 5.39 & 7.30  & 4.33 & 5.47\\
    & Picture description & 5.45 & 7.29 & 4.45  & 5.47\\
    \hline
    \end{tabular}
    \label{tab:performance_by_task}
\end{table}

\begin{figure*}
     \centering
     \begin{subfigure}[h]{0.3\textwidth}
         \centering
         \includegraphics[width=\textwidth]{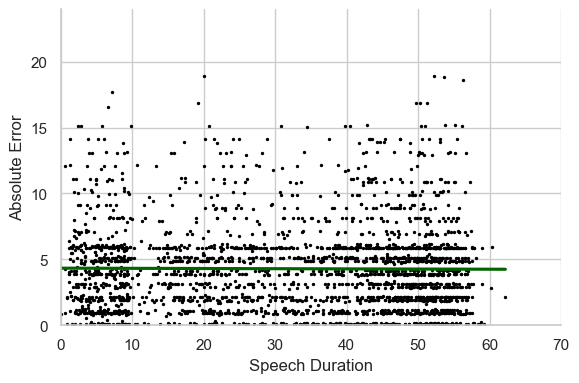}
         \caption{SVM (CCC = 0.000)}
         \label{fig:err_len_corr_svm_wll}
     \end{subfigure}
     \hfill
     \begin{subfigure}[h]{0.3\textwidth}
         \centering
         \includegraphics[width=\textwidth]{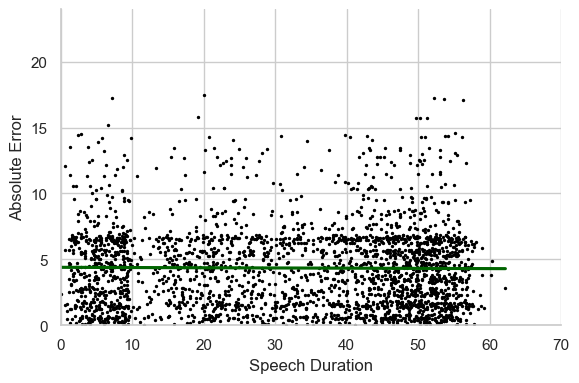}
         \caption{RF (CCC = 0.003)}
         \label{fig:err_len_corr_rf_wll}
     \end{subfigure}
     \hfill
     \begin{subfigure}[h]{0.3\textwidth}
         \centering
         \includegraphics[width=\textwidth]{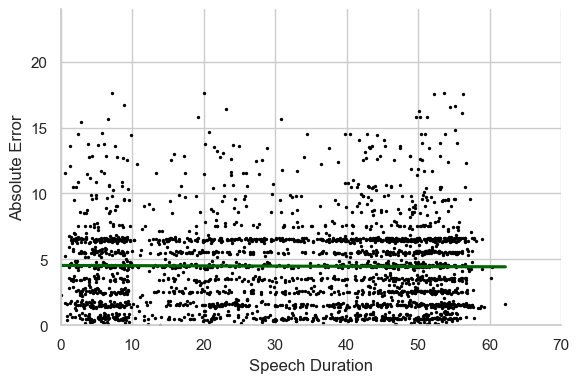}
         \caption{FNN (CCC = -0.002)}
         \label{fig:err_len_corr_fnn_wll}
     \end{subfigure}
        \caption{Correlation between speech length and prediction error of models trained on conventional acoustic features}
        \label{fig:err_len_corr_WLL}
\end{figure*}

\begin{figure*}[h]
     \centering
     \begin{subfigure}[h]{0.3\textwidth}
         \centering
         \includegraphics[width=\textwidth]{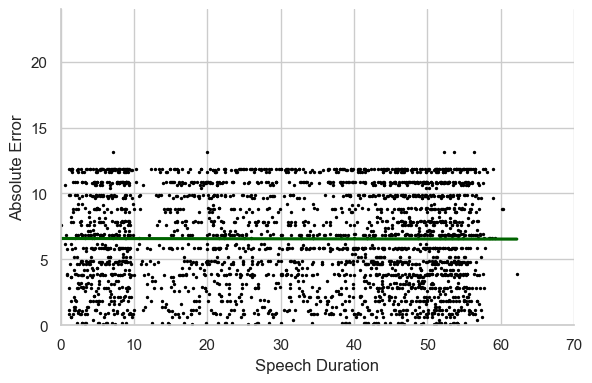}
         \caption{SVM (CCC = -0.004)}
         \label{fig:err_len_corr_svm_vgg}
     \end{subfigure}
     \hfill
     \begin{subfigure}[h]{0.3\textwidth}
         \centering
         \includegraphics[width=\textwidth]{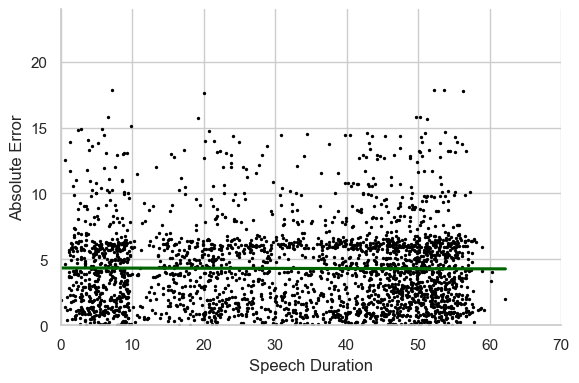}
         \caption{RF (CCC = -0.006)}
         \label{fig:err_len_corr_rf_vgg}
     \end{subfigure}
     \hfill
     \begin{subfigure}[h]{0.3\textwidth}
         \centering
         \includegraphics[width=\textwidth]{figure/VGG16/Err_len_corr_FNN.png}
         \caption{FNN (CCC = -0.010)}
         \label{fig:err_len_corr_fnn_vgg}
     \end{subfigure}
        \caption{Correlation between speech length and prediction error of models trained on VGG-16 features}
        \label{fig:err_len_corr_VGG}
\end{figure*}

\begin{figure*}[h]
     \centering
     \begin{subfigure}[h]{0.3\textwidth}
         \centering
         \includegraphics[width=\textwidth]{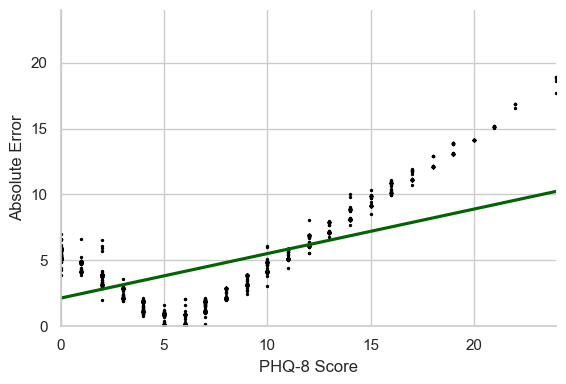}
         \caption{SVM (CCC = 0.055)}
         \label{fig:err_sev_corr_svm_wll}
     \end{subfigure}
     \hfill
     \begin{subfigure}[h]{0.3\textwidth}
         \centering
         \includegraphics[width=\textwidth]{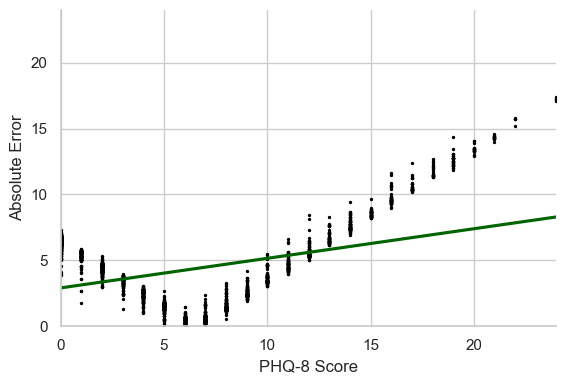}
         \caption{RF (CCC = 0.385)}
         \label{fig:err_sev_corr_rf_wll}
     \end{subfigure}
     \hfill
     \begin{subfigure}[h]{0.3\textwidth}
         \centering
         \includegraphics[width=\textwidth]{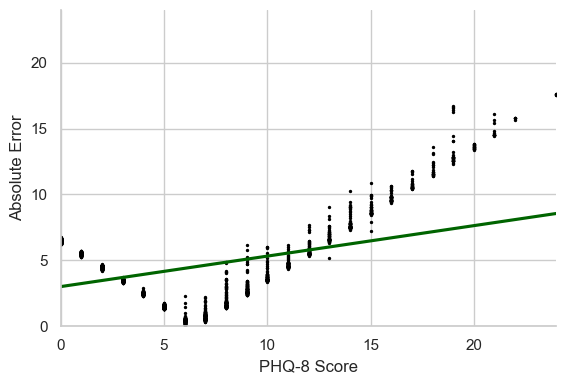}
         \caption{FNN (CCC = 0.435)}
         \label{fig:err_sev_corr_fnn_wll}
     \end{subfigure}
        \caption{Correlation between depression severity and prediction error of models trained on conventional acoustic features}
        \label{fig:err_sev_corr_WLL}
\end{figure*}

\begin{figure*}[h]
     \centering
     \begin{subfigure}[h]{0.3\textwidth}
         \centering
         \includegraphics[width=\textwidth]{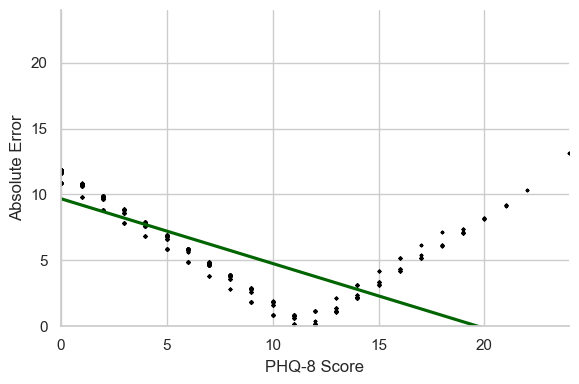}
         \caption{SVM (CCC = -0.726)}
         \label{fig:err_sev_corr_svm_vgg}
     \end{subfigure}
     \hfill
     \begin{subfigure}[h]{0.3\textwidth}
         \centering
         \includegraphics[width=\textwidth]{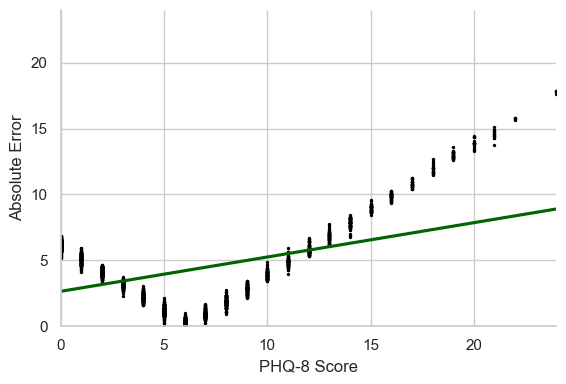}
         \caption{RF (CCC = 0.445)}
         \label{fig:err_sev_corr_rf_vgg}
     \end{subfigure}
     \hfill
     \begin{subfigure}[h]{0.3\textwidth}
         \centering
         \includegraphics[width=\textwidth]{figure/VGG16/Err_severity_corr_FNN.png}
         \caption{FNN (CCC = 0.394)}
         \label{fig:err_sev_corr_fnn_vgg}
     \end{subfigure}
        \caption{Correlation between depression severity and prediction error of models trained on VGG-16 features}
        \label{fig:err_sev_corr_vgg}
\end{figure*}

\section{Result and Discussion}
\label{sec: results} 
Here we present the performance of the ML models trained on DEPAC dataset with a view to find answers to the research questions (RQs), outlined in Section \ref{sec:intro}.
\subsection{Effectiveness of different types of acoustic features in measuring depression}
We trained 3 different ML models on each type of acoustic feature, i.e. conventional and VGG-16. We report the RMSE and MAE error of each model trained separately on samples from male and female subjects, along with the overall performance on the entire dataset (Table \ref{tab:overall_performance}). We compare the performance of our best model with the state-of-the-art (Table \ref{tab:comparison}). We report CPU time required to train each model to assist future researchers and system designers in making informed choice of feature type and ML model.

\subsubsection{Performance of models trained on conventional and deep representation features}
SVM and FNN models performed better on conventional features than on VGG-16, while performance of RF is marginally better (0.0004\%) on VGG-16 (Table \ref{tab:overall_performance}). These findings are consistent with the previous works presented in Table \ref{tab:comparison}. Conventional features presented in \cite{yang2020feature} modeled depression marginally better than models with deep representation features \cite{rodrigues2019multimodal}, \cite{zhao2020hierarchical}.

In comparison to the state of the art acoustic models, our proposed RF models show competitive performance. The RF model trained on both types of features outperforms almost all the existing works reporting similar performance metrics on PHQ-8 scale (see Table \ref{tab:comparison}). Only \cite{ray2019multi} reported lower RMSE than us, fusing all four sets of AVEC 2019 baseline features, which is a combination of conventional (MFCC, Bag-of-Audio Words, eGeMAPS) and deep representation (VGG-16) features, and formulating a multi-level LSTM architecture. Our proposed RF model trained on conventional features produces competitive performance to their proposed model, while substantially decreasing computational requirements. The VGG-16 features collected in the same manner as described by \cite{ray2019multi} from our audio corpus occupy 11.21 GB of memory, while our presented conventional feature file size is only 11 MB. Preprocessing and training models on conventional acoustic features took on average 3 minutes, while the procedure on VGG-16 features took at least 150 hours on the same computational environment (2.6 GHz 6 core Intel Core i7 processor, 16 GB memory). In short, our RF model using VGG-16 features offers 0.0004\% improvement in performance than the same model using conventional features, using 1000 times more memory and 3000 times more processing time, implying similar or more computational resource is required for training complex models on multimodal features for marginal performance improvement. Therefore, the conventional features provided better opportunity to adjust model parameters for performance improvement.

Results (Table \ref{tab:comparison}) demonstrate that, in most cases, RMSE and MAE are lower for male subjects than female subjects. The reason behind this can be the lower severity of depression among male subjects than females in DEPAC dataset \cite{tasnim2022DEPAC}. The skewness in the dataset causes bias in model prediction, as described in \cite{larrazabal2020gender}. For real world applications, this issue needs to be taken care of by ensuring gender balance in training data.

\subsubsection{Significance of performance deviation of models trained on conventional and deep representation features}

We performed two-sample t-tests to identify if the performance deviations of the models are significant when trained on conventional acoustic features and VGG-16 features.

There was a significant difference between absolute errors in predictions of SVM models trained on WLL acoustic features $(M = 4.28, SD =  3.25)$ and VGG-16 features $(M = 6.5, SD = 3.59); t(5332) = -24.20, p = 6.86e-123 < .05$. The absolute errors in predictions of SVM with model is significantly higher when trained on VGG-16 features than when trained on conventional features.

On the other hand, there was no significant difference between absolute errors in predictions of our best performing RF and FNN models trained on conventional acoustic features $( RF: M = 4.34, SD =  3.09; FNN: M = 4.32, SD =  3.14)$ and VGG-16 features $(RF: M = 4.31, SD =  3.11;  FNN: M = 4.48, SD = 3.11)$. The test scores of RF $(t(5332) = 0.38, p = .70 > .05)$ and FNN $( t(5332) = -1.81, p = .07 > .05)$ indicate that the deviation of errors in prediction of the models are not significantly different irrespective of training features. 

\subsection{Effect of speech task type on ML model performance}
The results (Table \ref{tab:performance_by_task}) do not reflect any significant deviation of model performance on the basis of speech task, therefore it is possible to recommend as a design choice any speech task of a similar length and content.

\subsection{Correlation between model performance and speech length}
From Figure \ref{fig:err_len_corr_WLL} and \ref{fig:err_len_corr_VGG} one can see that no significant correlation is observed between model performance (absolute error of each prediction) and length of corresponding sample. The Concordance Correlation Coefficient (CCC) score for SVM, RF and FNN models are 0.000, 0.003 and -0.002 for conventional features and -0.004, -0.006 and -0.010 for VGG-16 features respectively. The near-zero CCC values indicate that in the case of our dataset, the speech length of samples does not influence the models' performance. Note that all speech samples in DEPAC are less than one minute.

\subsection{Correlation between depression severity and model performance}
Absolute errors for each sample are plotted against ground truth PHQ-8 score for the models trained on conventional and VGG-16 features in Figure \ref{fig:err_sev_corr_WLL} and \ref{fig:err_sev_corr_vgg}. The CCC score for SVM, RF and FNN models are 0.055, 0.385 and 0.435 for conventional features and -0.726, 0.445 and 0.394 for VGG-16 features respectively. The plots, along with high positive CCC scores for most of the models, imply that the samples with higher PHQ-8 scores contribute more to the overall prediction error of the models. This is caused by the imbalance in the number of samples with high and low PHQ-8 scores in DEPAC dataset \cite{tasnim2022DEPAC}. The higher density of samples with subthreshold ($\leq$ 5) PHQ-8 score bias the models to make predictions close to the mean PHQ-8 (6.56) of the dataset. This observation strengthens the necessity of balancing the samples in training models to be used in real world application. 

\section{Conclusion and Future Works}
\label{sec: conclusion}
Speech has proven to be a reliable marker for depression assessment. But in order to deploy a machine learning model in a practical system, it is necessary to identify the most informative acoustic feature, along with an efficient and cost-effective process to train the model. In this paper, we study the performance of conventional acoustic feature-based and pre-trained deep representation based models on predicting depression severity from speech. We observe that the hand-curated feature based approach achieves better performance in terms of lower RMSE and MAE, at a remarkably less computation time. Our experiments show that gender of the speaker and distribution of score affect the model performance, and should be taken care of while formulating balanced training data. We also report that content and length of speech do not show significant impact as long as the length of speech samples is reasonably short, less than one minute in our case.  To summarize, we suggest using ML models trained on conventional features in resource-limited real-time situations and deep models in scenarios where fine-grained analysis involving higher computational power is crucial. In our future work, we plan to explore generalizability of the findings across other datasets and disorders.

\bibliographystyle{IEEEtran}
\bibliography{reference.bib}
\end{document}